%% file: main.tex
\documentclass[conference]{IEEEtran}

\IEEEoverridecommandlockouts

\usepackage{cite}
\usepackage{amsmath,amssymb,amsfonts}
\usepackage{algorithmic}
\usepackage{graphicx}
\usepackage{textcomp}
\usepackage{xcolor}
\renewcommand{\baselinestretch}{0.852}
\usepackage{tikz}
\usepackage{pgfplots}
\usepackage{pgfplotstable}
\pgfplotsset{compat=1.16}
\usetikzlibrary{patterns}
\usepackage{multirow}
\usepackage{enumerate}
\usepackage[caption=false]{subfig}
\usepackage[capitalise]{cleveref}
\usepackage[export]{adjustbox}
\usepackage{textcomp,mathcomp}
\usepackage{colortbl}
\usepackage{pifont}
\newcommand{\cmark}{\ding{51}}%
\newcommand{\xmark}{\ding{55}}%

\def\BibTeX{{\rm B\kern-.05em{\sc i\kern-.025em b}\kern-.08em
    T\kern-.1667em\lower.7ex\hbox{E}\kern-.125emX}}
\begin{document}

\title{Holistic IJTAG-based External and Internal Fault Monitoring in UAVs} 
\author{\IEEEauthorblockN{\large Foisal Ahmed, Maksim Jenihhin}
  \IEEEauthorblockA{Department of Computer Systems, Tallinn University of Technology\\
    Tallinn, Estonia\\}
  \IEEEauthorblockA{
    \{foisal.ahmed,maksim.jenihhin\}@taltech.ee } }

\maketitle

\input{Abstract}
\input{Introduction}

\input{Related_research_work_and_findings}
\input{Ijtag_based_health_monitor}
\input{Proposed_Methodology}
\input{Experimental_results}
\input{Conclusion}

\section*{Acknowledgments}
This work was partly supported by the European Union through the European Social Fund in the frames of the ‘‘Information and Communication Technologies (ICT) programme’’ (“ITA-IoIT” topic) and by the Estonian Research Council grant PUT PRG1467 “CRASHLESS".
\bibliographystyle{IEEEtran}

\end{document}

%% file: Abstract.tex
\begin{abstract}
Cyber-Physical Systems (CPSs), such as Unmanned Aerial Vehicles (UAVs), use System-on-Chip (SoC) based computing platforms to perform multiple complex tasks in safety-critical applications that require a highly dependable operation. Due to continuous technological manufacturing miniaturization SoCs face a wide spectrum of chip-level reliability issues such as aging, soft and hard errors during the operational lifetime of a UAV.  In addition, external (off-chip) faults in the sensors, actuators, and motors are another cause of UAV failures. While existing works examine either on-chip faults (internal) or sensors/actuators faults (external) separately, this research proposes a UAV health monitoring infrastructure considering both external and internal faults holistically. The proposed method relies on the IEEE 1687 standard (IJTAG) and employs on-chip embedded instruments as health monitors to instantly access external and internal sensor data. Experimental results for functional simulation of a real-life case-study design demonstrate both types of fault detection by serving only three clock cycles and the localization process using 16 and 30 clock cycles for the case of single and double faults, respectively.
\end{abstract}

\begin{IEEEkeywords}
UAV, IJTAG, IEEE 1687 Standard, Embedded Instruments, Health Monitoring, Dependability
\end{IEEEkeywords}

%% file: Introduction.tex
\section{Introduction}

Systems on Chip (SoCs), such as Multi-Processor System-on-Chip (MPSoC), Field-Programmable Gate Array SoC (FPGA-SoC), and recent complex All-Programmable SoC (APSoC) architectures, are becoming the standard computing platform for executing complex algorithms for real-time computation used in many Cyber-Physical System (CPS), edge-computing Internet-of-Things (IoT) devices, and Unmanned Aerial Vehicles (UAVs)~\cite{int1,foisal_survey}. Specifically, in mission and safety-critical applications,  the UAVs are engaged to search people who fall in distress or imminent danger, emergency public safety operations, inspection and maintenance of dangerous works, etc., demanding failure-free operations~\cite{SR5}. 


The dependability of these safety-critical applications is inherently related to the correct service of the UAV system which consists of several functional modules such as a flight control computer, communication module, global positioning system (GPS) module, neural network accelerator (NNA), etc. These functional modules are directly controlled and governed by the SoC located inside the UAV. Failures of UAVs may happen during the UAV operation such as position-altitude errors, crashes with obstacles, and target misidentifying due to the soft or hard errors in the SoC~\cite{SR64}.  Faults in external sensors such as sensors, actuators, and motors may also cause failure in the functional modules~\cite{SR56}.
Such scenarios may be even more catastrophic if failures of UAVs occur in mission- and safety-critical applications.

Careful focus on the reliability attribute is essential in designing the complex computing platforms of UAVs for fault-resilient operation in mission- and safety-critical applications. Most of the research works in UAV health monitoring cover various faults in external sensors, such as inertial measurement unit (IMU) sensors, actuators, GPS, battery~\cite{SR52, SR54, SR58, SR61},  and others focus on the soft or hard error, aging, process variations, etc., as internal (on-chip) faults in the UAV~\cite{SR63, SR64, SR66}. Health monitoring for external and internal defects in the complex UAV system is a challenging task when it considers the fault detection, isolation, and recovery in the different layers of the UAV system in a cross-layer manner~\cite{SR87}. During the lifetime operation of a SoC, IEEE 1687 (IJTAG)-based fault management infrastructure is a standard practice of health monitoring for the complex SoCs as described in~\cite{shibin2016, ali2019ijtag, ibrahim2019chip}.

In this paper, we propose an IJTAG-based UAV health monitoring by accessing External and Internal (referred to as Ex-In) sensor data. IJTAG standard provides an efficient and flexible way monitoring fault detection and managing fault recovery procedures during the system operation. While the existing methods consider either External or Internal sensor data, the proposed fault monitoring approach allows UAVs to monitor both Ex-In faults synchronously and exhaustively, which can help UAVs attain cross-layer fault adaptation and self-awareness. The feasibility of the proposed scheme is demonstrated by using a real-life example.

The rest of the paper is organized as follows. The related research work is briefly presented and the motivation for
the proposed work is discussed in Section II.  Section III  provides a background of the IJTAG-based health monitoring method. Section IV introduces the proposed design methodology. Section V presents the experimental setup, performed experiments, and the obtained results. Sections VI provides the conclusions of the paper.

%% file: Related_research_work_and_findings.tex
\section{Related Work}

In this section, we discuss faults and their effects on UAV systems addressed in recent research works. 
In~\cite{SR52, SR54}, the authors introduced a health management system based on Bayesian Networks (BNs) that monitors sensors and hardware components in real-time to detect and diagnose UAV failure. In their study, the authors examined UAV failures caused by the impact of GPS and battery consumption, as well as hardware (HW) and software (SW) failures resulting from weather disturbances. However, they did not account for on-chip soft and hard errors in their research.


As highlighted in~\cite{SR57}, actuators have a crucial function in translating control commands into actual control effects. However, actuator faults, such as being stuck, experiencing a partial loss of effectiveness, or having impairments in control surfaces, can result in mission failure and collisions among cooperative UAVs. In addition, a different research study examined actuator faults and an IMU fault, as reported in~\cite{SR58}. Through simulation, the researchers explored the impact of the actuator and gyroscope sensor faults. They reported that the roll, yaw rate, and side-slip angle were significantly affected due to the fault of the sensors and actuator. 


The authors discussed the occurrence of UAV position failures resulting from faults in UAV navigation sensors (such as GPS and IMU) in paper~\cite{SR61}. These faults primarily occur when the UAV operates in various environments. The authors of~\cite{SR62} presented two models to evaluate the system's overall reliability and investigate the consequences of error propagation in the inertial navigation system. These models assess the likelihood of various failures such as accelerometers, gyroscopes, temperature and pressure sensors, memory, GPS, etc., and also identify the critical components.



In~\cite{SR63, SR64}, the use of FPGA as a decision-making controller to regulate the accelerometer and motor of UAVs is discussed, with particular attention given to its susceptibility to transient (soft-error) and permanent fault (hard-error). These faults can lead to significant increases in vibration for the accelerometer and motor of the UAV. Additionally, a CNN accelerator based on FPGA is utilized for object identification and classification in the UAV. However, radiation-induced Single Event Upset (SEU) can cause bit flipping in the implemented CNN accelerator, resulting in incorrect results and misclassification rates.
In~\cite{SR66}, the authors conducted a study on the impact of radiation-induced errors on the SRAM-based FPGA that housed the neural network. The analysis of injecting faults, such as simulating SEU, in the FPGA-based MPSoC revealed that defects inside the CNN can affect overall accuracy, resulting in system failure, incorrect image categorization, and potential OS vulnerability. 

\begin{figure}[!t]
\centering
\includegraphics[width=.8\linewidth]{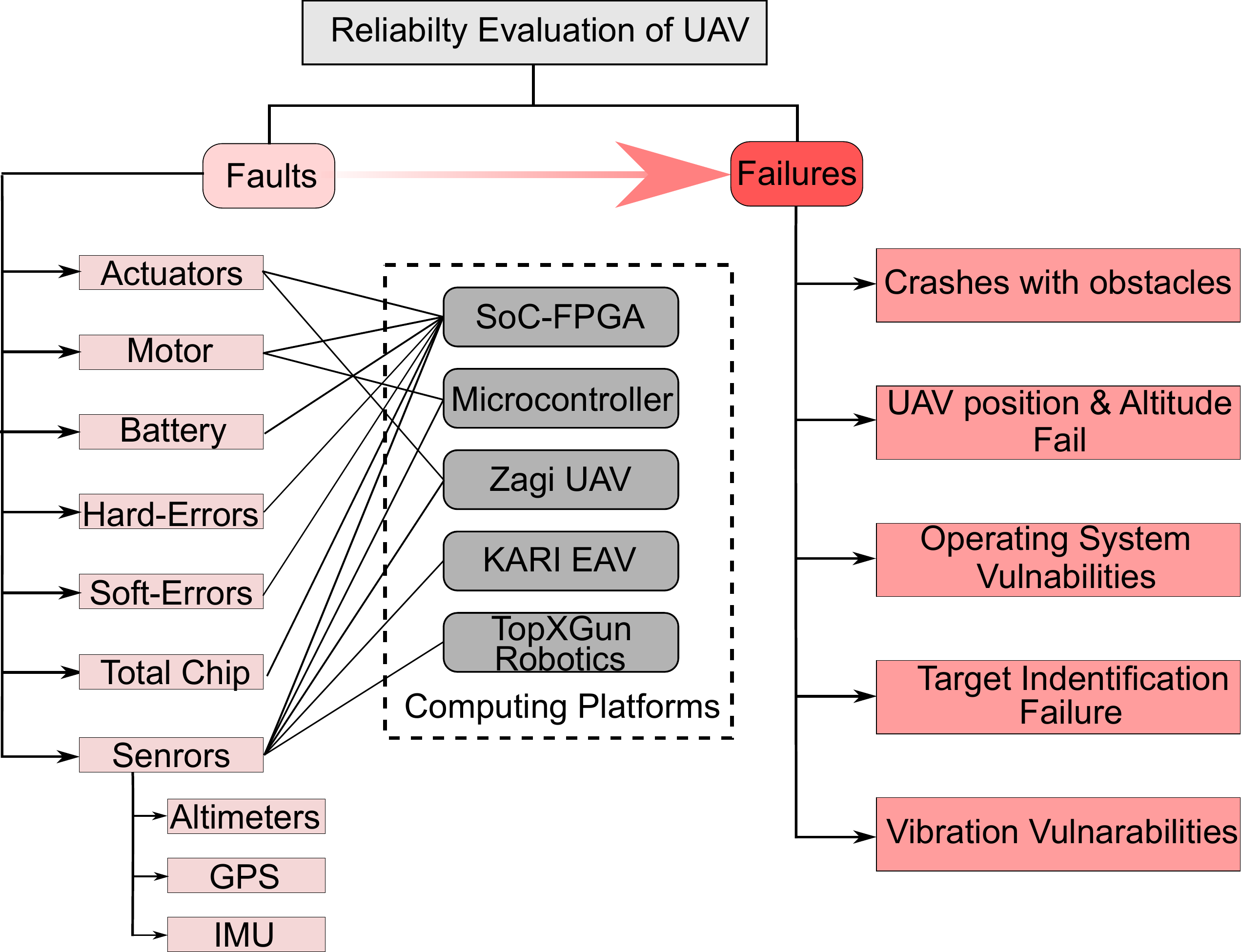}
\caption{UAV fault and failure mode analysis taxonomy~\cite{foisal_survey}.}
\label{fig3_new}
\end{figure}


Fig.~\ref{fig3_new} illustrates an evaluation of various UAV faults and failures that have been reported in recent studies~\cite{foisal_survey}. The majority of research work focuses on sensor faults such as actuators, motors, batteries, and GPS. Navigation faults in IMU sensors, such as accelerometers, gyroscopes, temperature, and pressure sensors, are also frequently discussed. Furthermore, soft-errors or transient faults are emerging as another significant cause of UAV system failures, particularly in the UAV SoC, as depicted in Fig.~\ref{fig3_new}.

As a result, considering the discussion above, we may broadly identify two types of faults. First, external (off-chip) faults happen in the IMU sensors, actuators, GPS, battery, etc. Second, soft, or hard error, aging, process variations, etc., as internal (on-chip) faults in the SoC also cause the failure of the UAV. The main object of this work is to manage these external and internal faults for UAV health monitoring holistically.

It should be noted that, while malicious faults are an important concern for drones serving for mission- and safety-critical applications, security aspects are out of scope of this paper.

%% file: Ijtag_based_health_monitor.tex
\section{Overview of IJTAG-Based Health Management}

The main requirements for any on-chip fault management should be faster error detection capabilities, high fault-localization ability, reliability, and scalability of the service infrastructure~\cite{shibin2016}. 

\begin{figure}[!tbp]
  \centering
  \begin{minipage}[b]{0.2\textwidth}
    \includegraphics[width=\textwidth]{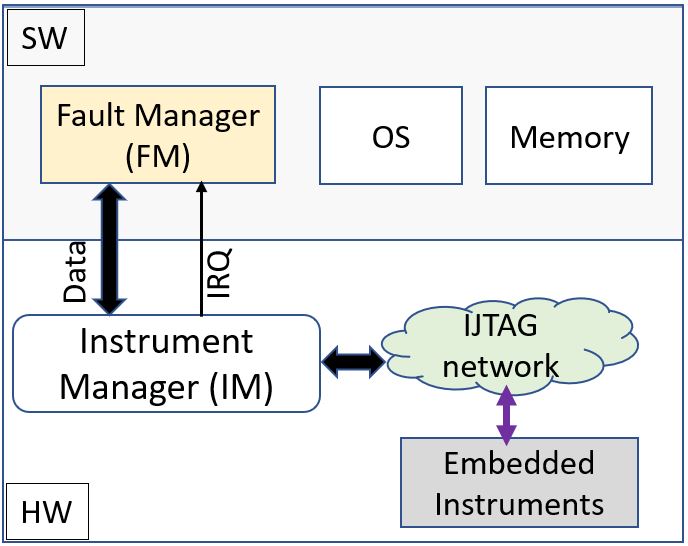}
    \caption{Basic concept of fault management infrastructure~\cite{shibin2016}.}
    \label{fig:HM_IJTAG}
  \end{minipage}
  \hfill
  \begin{minipage}[b]{0.23\textwidth}
    \includegraphics[width=\textwidth]{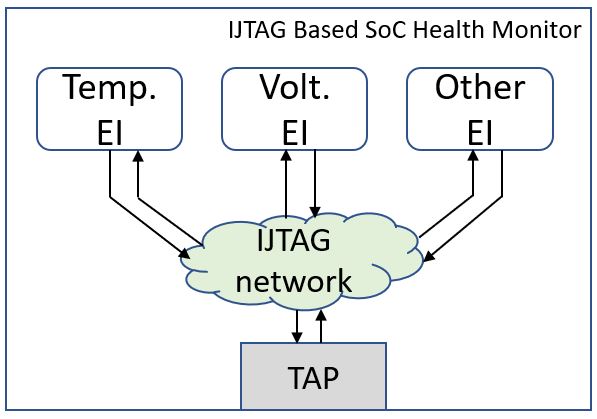}
    \caption{Overview of on-chip Health Monitors~\cite{ali2019ijtag}.}
    \label{fig:SoC_HM}
  \end{minipage}
\end{figure}


In this section, we discuss the different fault management scenarios and on-chip health monitoring systems used for safety and reliability applications~\cite{shibin2016, ali2019ijtag, ibrahim2019chip}.


\begin{figure*}[!tbp]
  \centering
  \subfloat[The proposed design for accessing Ex-In sensor data.]{\includegraphics[width=0.67\textwidth]{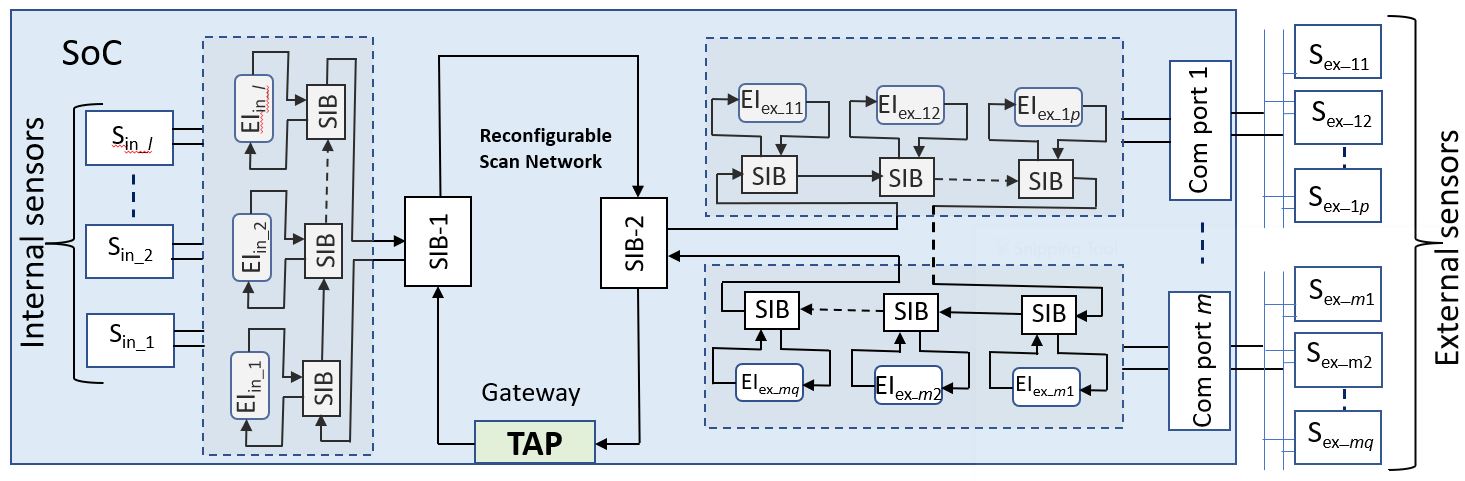}\label{fig:Proposed}}
  \hfill
  \subfloat[Basic concept of fault management of the proposed method.]{\includegraphics[width=0.3\textwidth]{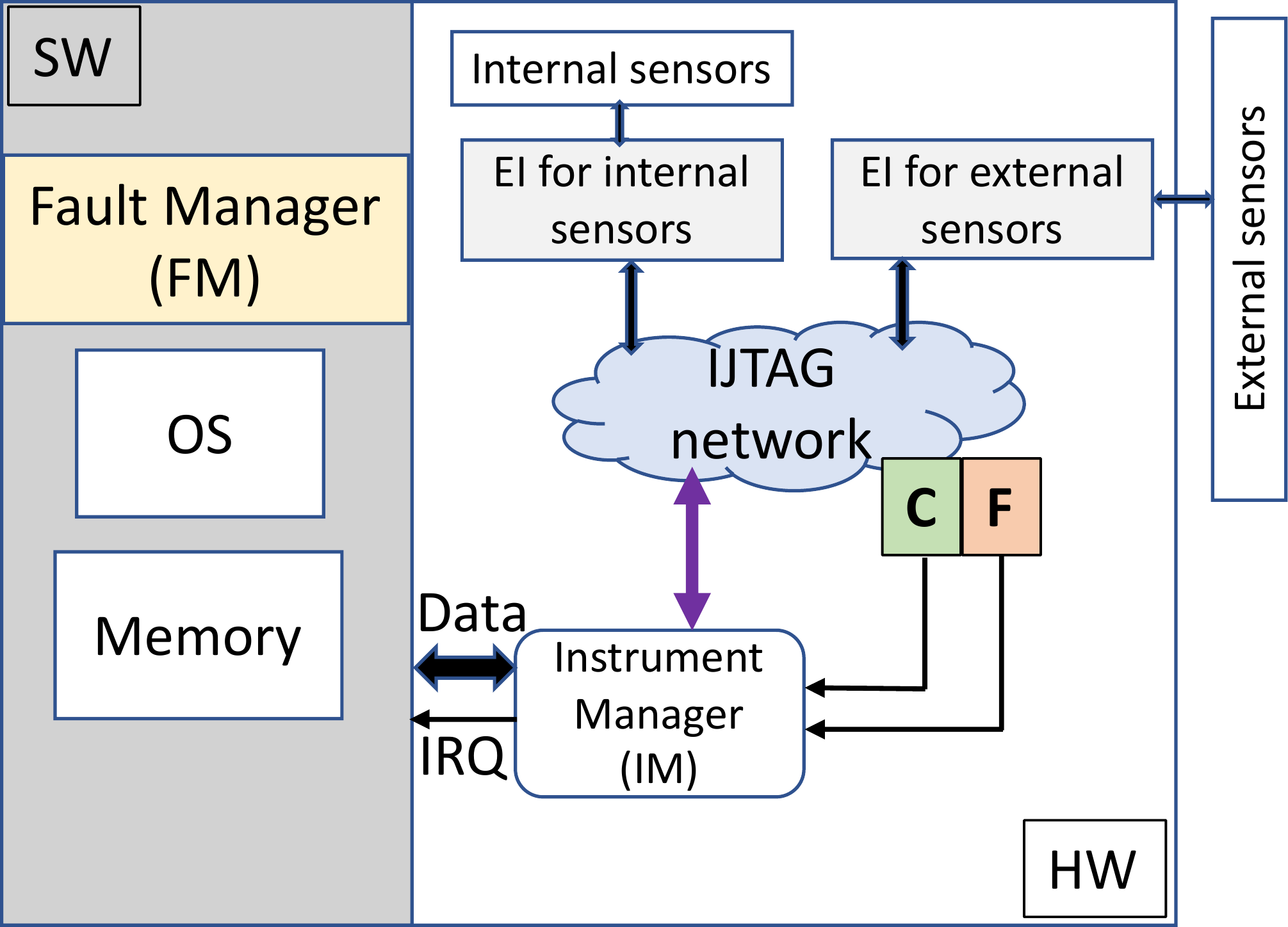}\label{fig:hm_proposed}}
  \caption{Proposed IJTAG-based fault monitoring for UAV.}
\end{figure*}

The first idea of an IJTAG-based fault management system for complex SoCs was proposed in~\cite{shibin2016} where the authors utilized IJTAG in the cross-layer fault management methodology on the system-wide integration including many components such as service HW, middleware, OS, and application SW. ~\cref{fig:HM_IJTAG} depicts the general concept of the fault management framework where the fault manager (FM) in the SW and instrument manager (IM) in the HW are tightly coupled with the functional part of the system. The resource maps and system health are governed by FM, which serves as a service management SW. FM and IM share health-related information. IM is an implemented HW component directly coupled to the IJTAG infrastructure to monitor on-chip embedded instruments (EIs). The EIs are connected to different or similar IP cores, which are prone to deterioration or defect. For graceful degradation, these IP cores should be able to replace one another totally or partially.



We have also investigated other research works~\cite{ali2019ijtag, ibrahim2019chip} for IJTAG-based fault management where they considered the functional, safety, and reliability applications. The authors in work~\cite{ali2019ijtag} deployed in-situ EIs for monitoring online timing, voltage, and temperature to ensure dependable operation during its operational lifetime. All EIs in the health monitors (HM) are IJTAG compatible. ~\cref{fig:SoC_HM} shows the overview of their proposed HM system. They observed the voltage and temperature variations by the Temperature and Voltage EI, respectively. As the propagation delay of the critical path inherently depends on the process variations, operating voltage, and temperature, so they designed timing EI in such a way that the exact remaining slack can be measured despite the voltage and temperature variations. These EIs are configured by IJTAG so that data for all sensors are accessed and captured at the same time. Finally, an embedded processor is used to fuse the data obtained from EIs to estimate the aging and remaining lifetime.

%% file: proposed_methodology.tex
\section{Proposed Methodology}

Currently, most of the fault monitoring methods for UAVs consider either sensors/actuators faults denoted as external faults (EF) or faults that occurred in the computing platforms denoted as internal faults (IF) discussed in Section II. The main objective of this research work is to monitor both EF and IF faults by observing Ex-In sensor data through IJTAG-based on-chip EIs. ~\cref{fig:Proposed} shows the basic concept of our proposed design using two branches of EIs for monitoring IF and EF, respectively. The proposed design uses a few sets of EIs as an example for accessing the Ex-In sensor. The accurate number of EIs relies on dependable applications and the complex structure of SoC.

Firstly, the internal sensors $(\text{S}_{\text{in}\_1},\text{S}_{\text{in}\_2}, ..., \text{S}_{\text{in}\_l})$ are used to observe on-chip health information such as temperature, voltage, delay, etc. and these data are captured by using a set of EIs as $(\text{EI}_{\text{in}\_1},\text{EI}_{\text{in}\_2}, ..., \text{EI}_{\text{in}\_l})$   (left-side in~\cref{fig:Proposed}), where $l$ is the number of internal sensors. Due to the variation of temperature, voltage, and process corner, the propagation delay of a critical path changes from the nominal value or causes the aging problem. Using a ring oscillator as an on-chip EI can be used for measuring the delay information as aging~\cite{8872109}. In addition, the aging effects due to temperature and process variations during operation also increase the soft-error rate (SER) considerably~\cite{cannon2008impact} and make misclassification problems in the CNN-based target identification and decision-making controller~\cite{SR66}. Variations in voltage, temperature, and delay can be examined using EIs as~\cite{ali2019ijtag}.

In the proposed design, the external sensor data can be captured through the $m$ number of functional ports (Com port 1, Com port 2, ..., Com port $m$) via standard serial communication protocol (e.g. UART, I2C, SPI, etc.)~\cite{com}. Similarly, other sets of EIs $(\text{EI}_{\text{ex}\_11},\text{EI}_{\text{ex}\_12}, ..., \text{EI}_{\text{ex}\_1p})$ and $(\text{EI}_{\text{ex}\_m1},\text{EI}_{\text{ex}\_m2}, ..., \text{EI}_{\text{ex}\_mq})$ in the proposed design (right-side in~\cref{fig:Proposed}) are employed for accessing the health information of external sensors $(\text{S}_{\text{ex}\_11},\text{S}_{\text{ex}\_12}, ..., \text{S}_{\text{ex}\_1p})$ and $(\text{S}_{\text{ex}\_m1},\text{S}_{\text{ex}\_m2}, ..., \text{S}_{\text{ex}\_mq})$, respectively, such as IMU, GPS, Battery, etc. Here, $p$ and $q$ are the number of external sensors accessed through Com port 1 and Com port $m$, respectively. Faults in these external sensors cause mission failure when the UAV is operating in a different environment~\cite{SR61}. These EIs can access the external sensor data to monitor EF information~\cite{SR62}.

The accessed data from the EIs can be further analyzed using the embedded IM to detect the fault accurately and manage the fault-recovery procedure during the operation~\cite{shibin2016}. IM uses the asynchronous fault propagation network where the health information in the Ex-In sensors is immediately reported to the CPU by the flag-based error propagation system. Then, the IM sends an interrupt signal to the CPU to stop the running operations of the UAV and starts the localization procedure of the faulty sensor using the flag information. Based on the fault information, SW running in the CPU can classify the fault and update the system health information of the UAV.~\cref{fig:hm_proposed} shows the basic concept of the fault management system used in our proposed method to handle the faulty Ex-In sensors. The proposed design employs the extended SIB (Segment Insertion Bits) in the IJTAG network which includes three additional flag registers for faulty (F), correction (C), and mask (X) signals in the asynchronous fault detection system~\cite{jutman2016}. The main advantages of the proposed method are that EF and IF are observed synchronously and immediately by configuring the EIs through the IJTAG network and handling both Ex-In faults at the same time to achieve failure-free UAV operation. The proposed design achieves cross-layer adaptation by monitoring internal and external status of the UAV computing platform. Thus, it enables an efficient fault-tolerance and self-awareness of  UAVs. 


The proposed design is discussed in the next subsection with two different case studies.

\subsection{External and Internal Sensor Access}
It is not a trivial way to determine the exact number, location, and types of EIs. Generally, it depends on the applications and quality of service. 


\subsubsection{Case-Study-1: XADC as internal Sensors}
\begin{figure}[h!]
\includegraphics[width=0.7\linewidth]{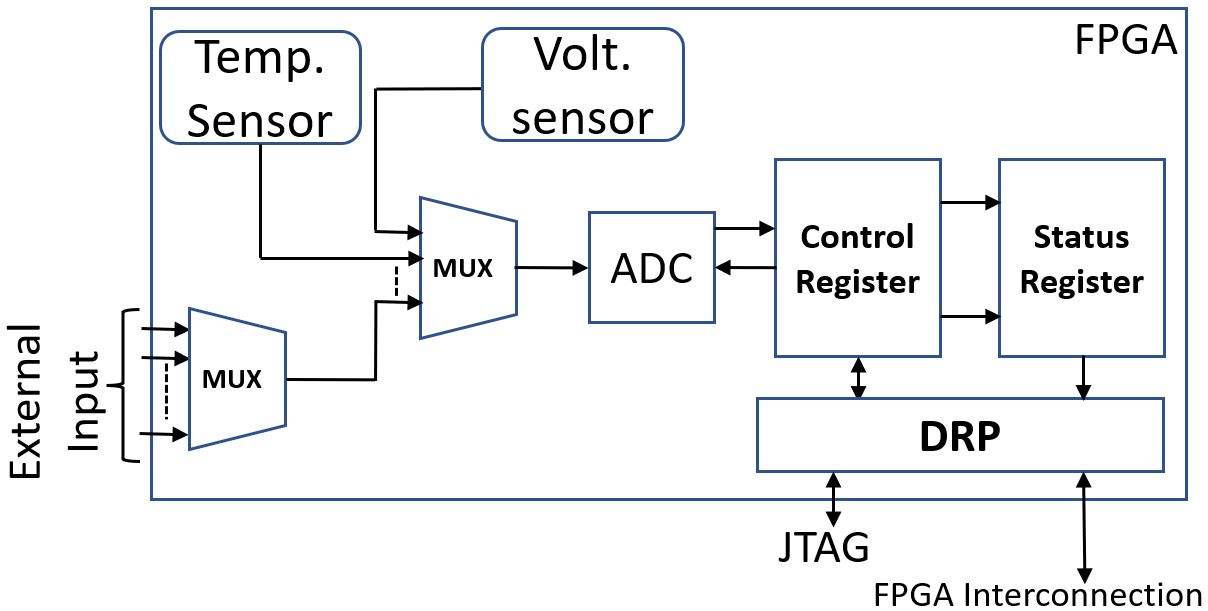}
\centering
\caption{XADC block diagram\cite{xadc}.}
\label{fig:xadc}
\end{figure}

\begin{figure}[h!]
\includegraphics[width=0.9\linewidth]{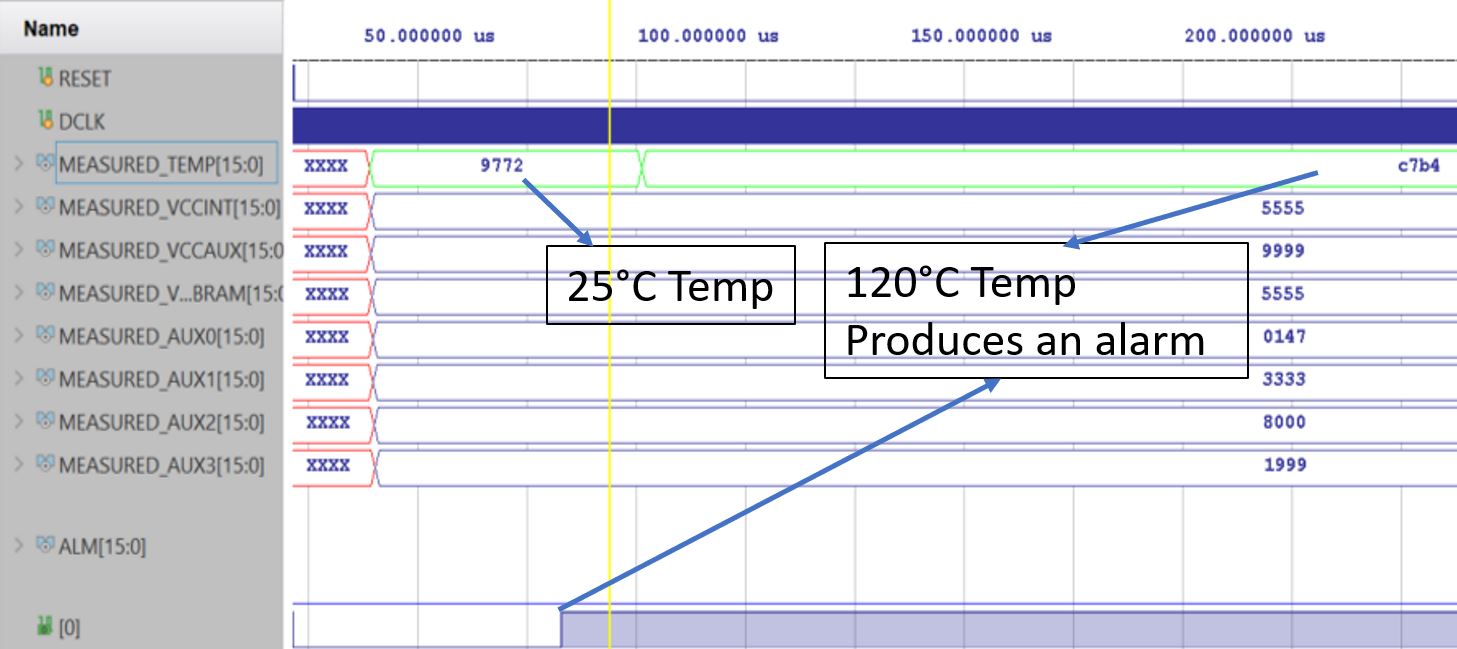}
\centering
\caption{XADC simulation results for temperature and source voltages.}
\label{fig:xadc_result}
\end{figure}
In this study, we have used the Xilinx analog to digital conversion (XADC) module~\cite{xadc} to monitor on-chip temperature and voltage. Xilinx XADC is a popular module used in many applications such as Data acquisition systems, IoT monitoring, and conversion of analog sensor data into digital~\cite{moroz2019application}. ~\cref{fig:xadc} shows the basic block diagram of the XADC controller module  that includes internal on-chip temperature and voltage sensors to measure die-temperature and power supply voltages, ADC module that converts data coming from both internal sensors and external analog input, and memory registers. The status register stores the converted data accessible by FPGA interconnection or JTAG port through Dynamic Reconfigure Port (DRP) shown in~\cref{fig:xadc}. There are also control registers to configure the XADC operations as XADC alarm functions, factory tests, etc. XADC also produces an alarm signal when the internal sensor measurements exceed a threshold value.  

To observe the XADC operation, an analog stimulus file was used as testbench analog signals (Temperature and Voltages). The output of a simple behavioral simulation is shown in~\cref{fig:xadc_result} to observe the sensor data in the status register. From this figure, we can observe the digital data after the conversion of analog stimulus signals of die temperature and source voltages. E.g., in the testbench stimulus file, we stored analog temperature values of 25$\tccentigrade$ and 120$\tccentigrade$ at a different interval. XADC module first converts the on-chip sensors data into digital and stores them in the status registers as shown in the~\cref{fig:xadc_result}. We can also notice that an alarm signal rises when the temperature exceeds the threshold value.

\subsubsection{Case-Study-2: IMU as External Sensors}
\begin{figure}[t!]
\includegraphics[width=0.7\linewidth]{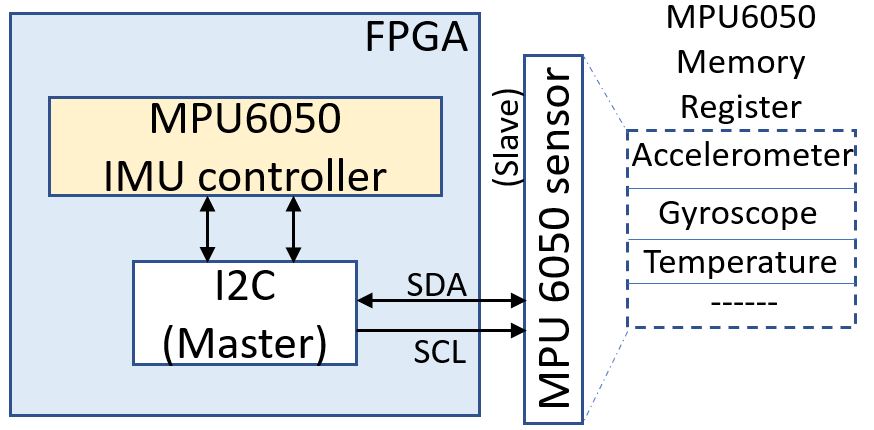}
\centering
\caption{Block diagram of MPU6050.}
\label{fig:mpu6050}
\end{figure}
\begin{figure}[t!]
\includegraphics[width=\linewidth]{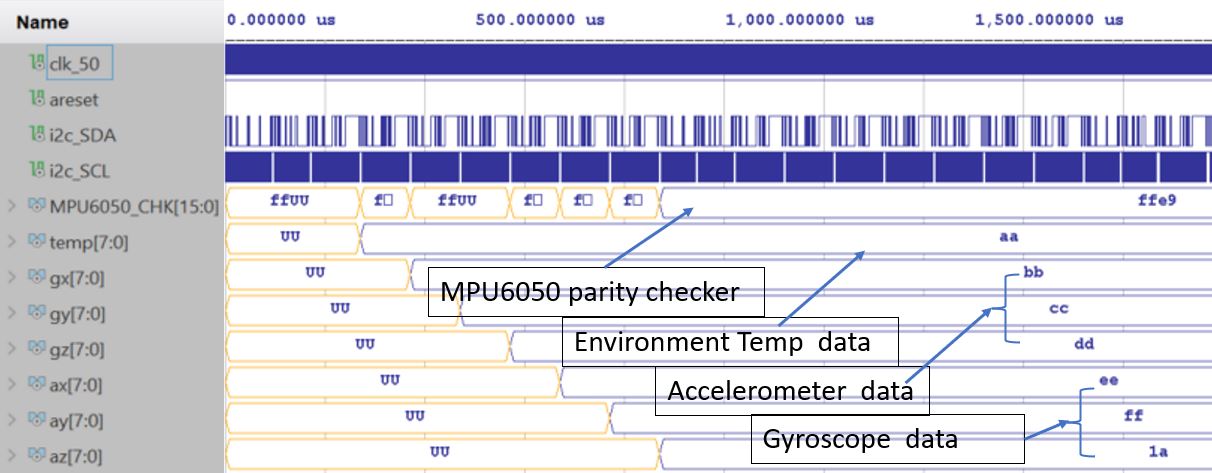}
\centering
\caption{Simulation result of the MPU6050 sensor module.}
\label{fig:mpu6050_result}
\end{figure}
As we have seen in Section II, the sensors such as accelerometers, gyroscopes, and GPS are important parts of the UAV system. The correct operations like tracking, object detection, and localization depend on the fault-free performance of those sensors. In this work, we have used an IMU sensor e.g., MPU6050 sensor module in our case-study as~\cite{kumari2017interfacing}. I2C COM port is created and connected to communicate with MPU6050 sensor and designed MPU6050 controller in the FPGA shown in~\cref{fig:mpu6050}. The sensor data is updated and stored continuously in the status register according to the UAV's orientation during flight operation. A parity checker is also created in the MPU6050 controller unit to check the IMU sensor data. Sample raw data was used on the memory file of the MPU6050 sensor module for the simulation purpose. ~\cref{fig:mpu6050_result} demonstrates the simulation results to observe the external sensor data of the accelerometer, gyroscope, and temperature. We can also monitor the MPU6050 parity checker data according to the sensor raw data.

%% file: Experimental_results.tex
\section{Experimental Results}
\begin{figure}[t!]
\includegraphics[width=0.88\linewidth]{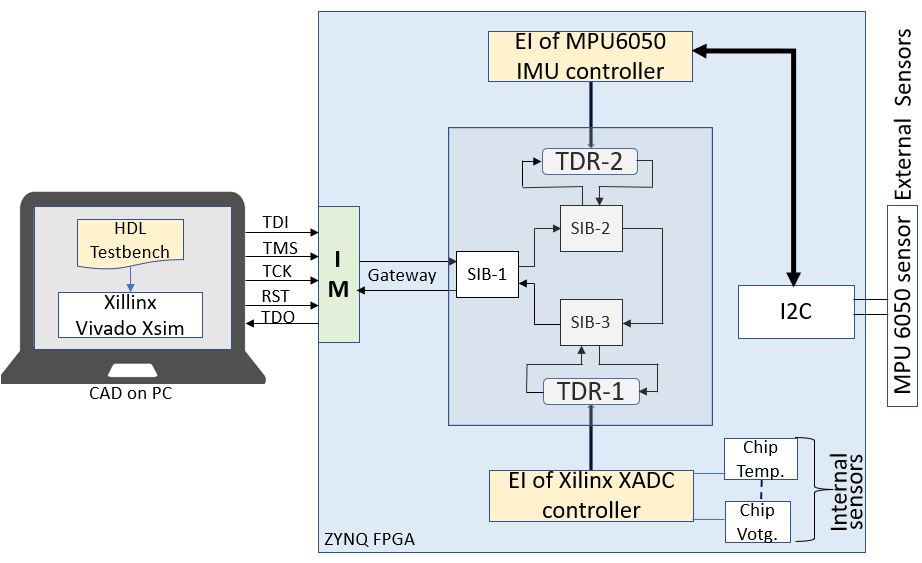}
\centering
\caption{Experimental setup for the evaluation.}
\label{fig:ex_setup}
\end{figure}

\subsection{Experimental Setup}
To evaluate our proposed method, the Xilinx ZC702 evaluation board (XC7Z020dg484-1) for the ZYNQ APSoC was used in this experiment.~\cref{fig:ex_setup} shows the experimental setup where we implemented an IJTAG network with 3 SIBs and 2 TDRs (Test Data Registers) to access the Ex-In sensor data. The controllers for IMU and I2C master-slave were designed for accessing external MPU6050 sensor data. The output of the IMU checker register is connected with one of the TDRs of the IJTAG network. For the internal sensor, the on-chip sensor of temperature and voltage from the Xilinx XADC core was also included in another TDR of the IJTAG network. The IM was also implemented to handle fault management procedures in this network. The address of the SIBs and TDRs is mapped in the ROM file for the IM. To check the functionality of the proposed design, an HDL testbench was created and conducted the functional simulation in the Vivado Xsim simulator environment. 


\subsection{Result Discussion}
\begin{figure}[h!]
\includegraphics[width=\linewidth]{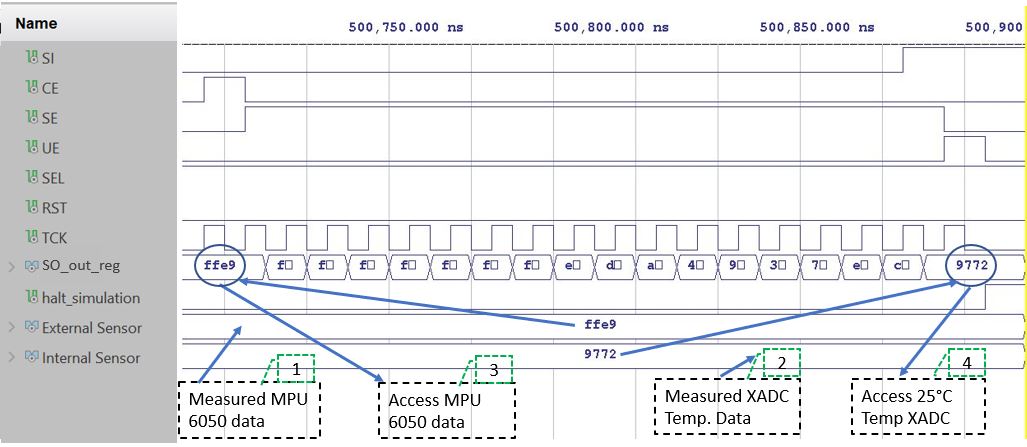}
\centering
\caption{Functional simulation results of our proposed design.}
\label{fig:final_sim}
\end{figure}
First, only two SIBs with IJTAG networks without the IM were used to observe both Ex-In sensor data.~\cref{fig:final_sim} presents the system-level functional simulation results. The simulation results show only the access and read operation of the IMU checker and on-chip temperature value for the external and internal sensors, respectively. Other sensor values could also be shown correspondingly. The test vectors in the testbench configure the SIB-3 and SIB-2, respectively, to read the TDR's value and shift-out the value to the output SO. Important observations in~\cref{fig:final_sim} are highlighted as \(\text{``$1$"}\), \(\text{``$2$"}\), \(\text{``$3$"}\), and \(\text{``$4$"}\).


After some delay (not shown in the figure), the measured data of the MPU6050 sensor and the on-chip temperature of XADC were available to access using the IJTAG network (1)(2). Then at first, SIB-2 was configured by loading \(\text{`$01$'}\) to make the scan chain with TDR-2. After the shifted corresponding number of bits, the data of the MPU6050 sensor was observed at the SO output register (3) as a similar result was found in the~\cref{fig:mpu6050_result}. Finally, we configured SIB-3 by loading \(\text{`$10$'}\) into the TDR-1 register which is parallelly connected with the XADC module. Similarly, we shifted the zero values to monitor the on-chip temperature of 25$\tccentigrade$ (4) as compared with the previous XADC temperature result shown in~\cref{fig:xadc_result}. Thus, the output result of the proposed design in~\cref{fig:final_sim} revealed that both Ex-In sensors data are accessed synchronously and immediately.
\begin{figure}[t!]
\includegraphics[width=\linewidth]{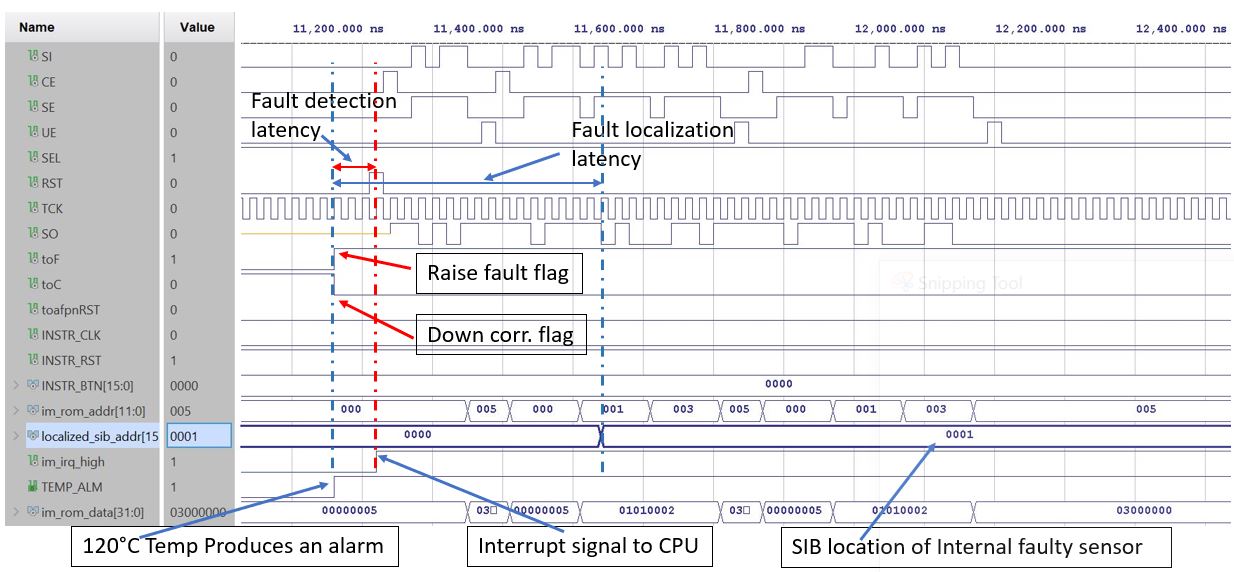}
\centering
\caption{ Result of single fault reporting and localization for the internal sensor.}
\label{im_result_in}
\end{figure}


\begin{figure}[h!]
\includegraphics[width=\linewidth]{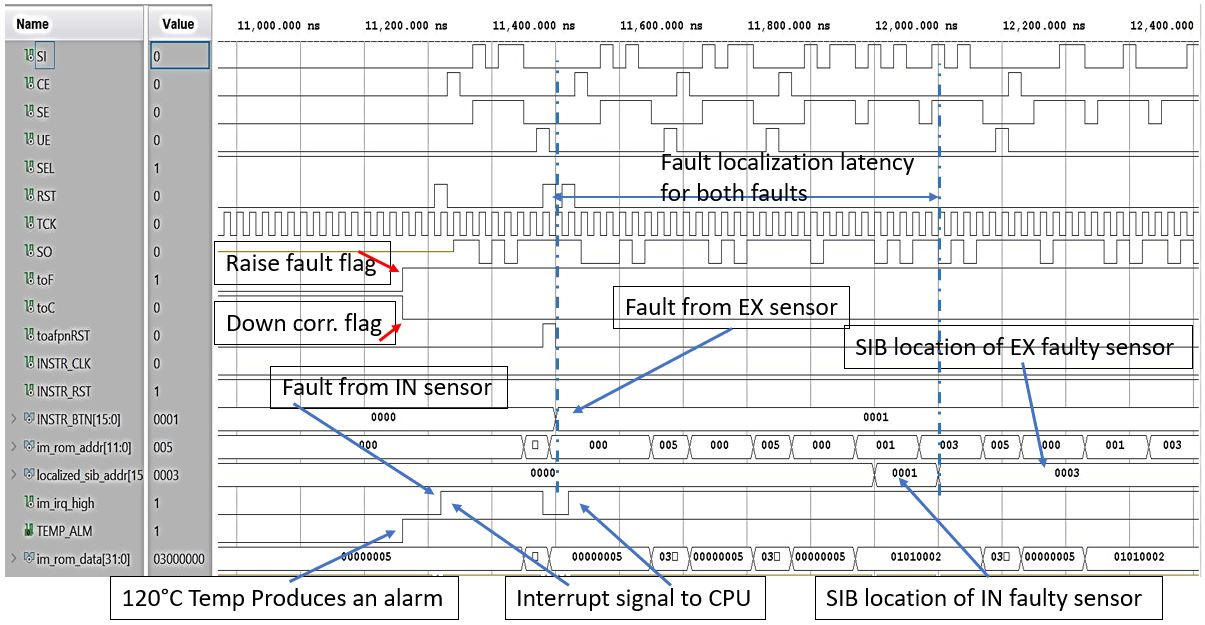}
\centering
\caption{Result of multiple faults reporting and localization for both Ex-In sensors.}
\label{im_result_exin}
\end{figure}

To observe the fault management procedures of the proposed method, an IM was included with another SIB (SIB-1) as shown in~\cref{fig:ex_setup}. Before the operation, all SIBs and TDRs are mapped based on the IJTAG network where a 16-bits address was used for each location in the ROM file. E.g. the address of SIB-1, SIB-2, and SIB-3 are \(\text{`$0000$'}\), \(\text{`$0003$'}\), and \(\text{`$0001$'}\), respectively in the hexadecimal format. Both single and multiple faults were analyzed in the experiment.~\cref{im_result_in} shows the fault management scenario due to the internal sensor fault, in this case, an alarm signal rises when the temperature exceeds the threshold value of 120$\tccentigrade$ for instance as discussed in the previous section shown in~\cref{fig:xadc_result}. From this Figure, we can observe that as soon as the alarm signal rises, the F and C in the IJTAG network are changed instantly and IM sends an interrupt signal after three cycles which can be used for the CPU to stop the current operation. After then IM starts the fault localization procedures which take a total of 16 cycles to localize the address. It is clearly observed from the Figure that IM localizes the SIB-3 address \(\text{`$0001$'}\) of the internal sensor when raised an alarm signal.

\begin{table}[!h]
    \centering
         \fontsize{8}{8}\selectfont
     \caption{\label{tab-1}The latency for different fault handling cases}
    \begin{tabular}{ccc}
    \hline
        \textbf{Latency} & \textbf{Time, clock cycles} & \textbf{Time, $\mu$s at 200\,MHz} \\
        \hline\hline\\
         Detection latency & 3 & 0.015 \\ 
         for single fault  &  &  \\ 
         Localization      & 16 & 0.08 \\ 
         for single fault  &  &  \\ 
         Localization      & 30 & 0.15 \\ 
         for two faults    &    &  \\ 
     \hline
    \end{tabular}
\end{table}

The proposed method was also analyzed for multiple fault scenarios shown in~\cref{im_result_exin}. From this figure, we can also observe that IM can detect and locate multiple faults similar way. In this case, after detecting and locating the internal faults, the IJTAG network was reset and then a fault was injected by a button switch externally to mimic the fault in the external sensor. The IM also detected and localized both faults by using 30 cycles shown in Figure. It first localized the SIB-3 (\(\text{`$0001$'}\)) and SIB-2 (\(\text{`$0003$'}\)) for the internal and external sensors, respectively. Table~\ref{tab-1} shows the latency required for different fault detection and localization cases. The logic resources utilization for this tiny IJATG network with IM on the ZYNQ FPGA-based implementation were only 0.48\% LUTs and 0.31\% FFs. As an important result of the proposed IJTAG-based UAV fault detection method, we can observe a single fault detection (external or internal) using only 3 cycles as all the EIs are connected in the same IJTAG network which is used as the standard practiced industrial approach.


\begin{figure}[h!]
\includegraphics[width=\linewidth]{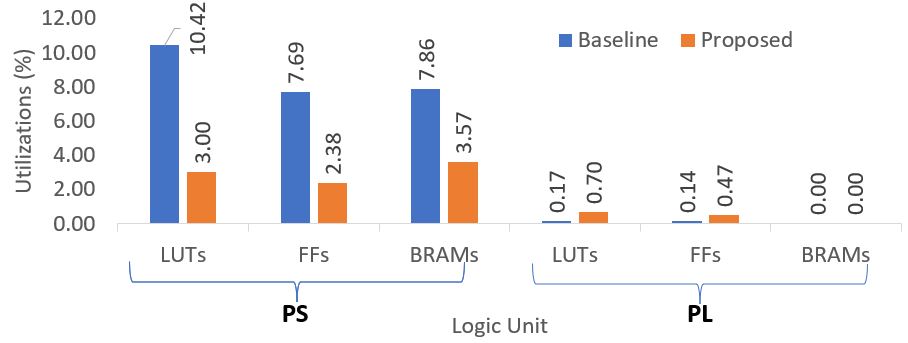}
\centering
\caption{Comparison of the proposed method with the baseline system for resource utilization.}
\label{fig:comapre}
\end{figure}    
We compared the proposed approach with the baseline system without IJTAG-based on-chip health monitoring for the two Ex-In instruments. To explore the design space of both implemented systems, a ZYNQ processing system (PS) based on ARM Cortex-A9 core was connected to the programmable logic (PL) unit by using a stream-based advanced extensible interface (AXI)~\cite{zynq}. The direct memory access (DMA) controller was included with the PS to store and access the data of the instruments in the external memory. The~\cref{fig:comapre} represents the resource utilization used for both systems. It is observed clearly from the Figure that the proposed design requires less logic utilization in all cases than the baseline system in PS implementation. For the LUTs and FFs, the proposed technique achieves around 71.20\% and 69\% less logic utilization, respectively. The baseline system requires an additional integrated logic analyzer (ILA) to monitor the internal signal~\cite{ila} and two separate DMA controllers for both Ex-In instruments. 

On the other hand, the proposed method needs a few more logic resources than the baseline system in the PL scenario. The reason is due to the on-chip fault monitoring system connected to the IJTAG network. However, in overall SoC implementation, the proposed approach reduces 65\% LUTs, 63.87\% FFs, and 54.58\% BRAMs utilization, respectively. The main advantage of this proposed design over the baseline is it can be easily scalable. The modern SoC requires thousands of EIs, which can be quickly added to this proposed system without increasing the PS resources significantly. However, as the number of EIs grows, the resource in PS of the baseline system tends to increase notably.


Finally, we compare our proposed design with the related works~\cite{ali2019ijtag, SR56, SR64} in Table~\ref{tab-3}. All works in the table used FPGA-based implementation for two modules as the case studies considering external or internal sensors. Our proposed work considers both Ex-In fault monitoring to achieve the goal of cross-layer health management holistically. The work in~\cite{ali2019ijtag} also used a similar IJTAG-based fault management infrastructure that included only the internal sensors for monitoring the voltage and temperature. The works~\cite{SR56, SR64} did not consider both Ex-In faults and they only focused on the fault detection, not localization steps.

\begin{table}[!t]
    \centering
         \fontsize{8}{8}\selectfont
     \caption{\label{tab-3}The comparison of the proposed method}
    \begin{tabular}{| m{3em} | m{0.7cm}| m{0.7cm} | m{0.7cm} | m{6em}| m{0.8cm} |m{0.6cm} | }
    \hline
        \textbf{Method} & \textbf{In.} & \textbf{Ex.} & \textbf{FDL} & \textbf{Faults} & \textbf{Method}& \textbf{Used} \\
              & \textbf{Sensor} & \textbf{Sensor} &  &  &  &\\ 
        \hline\hline
        Ref~\cite{SR64} & \cmark & \xmark & \xmark & SEU & Markov & UAV \\ 
         Ref~\cite{SR56} & \xmark & \cmark & \xmark & Battery \& GPS & Bayesian & UAV \\ 
         Ref~\cite{ali2019ijtag} &\cmark  & \xmark  & \xmark & Volt. \& Temp. & IJTAG & Car \\ 
        Proposed & \cmark & \cmark & \cmark & IMU, Volt.,\& Temp. & IJTAG & UAV \\  \hline
        \multicolumn{7}{|c|}{FDL=Fault detection \& localization} \\ \hline
    \end{tabular}
\end{table}

%% file: Conclusion.tex
\section{Conclusion}

In this paper, we presented a novel Ex-In sensor access methodology for the UAV self-health awareness infrastructure based on IEEE 1687 standard. Embedded instruments in the proposed design simultaneously monitor on-chip variations such as temperature, voltages, delay, soft- and hard errors, etc., and off-chip faults in external sensor such as IMU, GPS, etc. Simulation results with real-life scenarios show that the proposed approach can access both Ex-In health information to detect and localize external and internal faults ensuring dependable UAV operation synchronously and immediately. Only three cycles are required for detecting any Ex-In fault where the localization latency is only 16 cycles after occurring the single fault. The proposed design achieves more than 50\% less logic utilization than the baseline system for implementing the UAV SoC-health.   
